\documentclass[a4paper, twoside, reqno, 12pt, dvips]{amsart}
\RequirePackage{fix-cm} 

\usepackage{fixltx2e}     

\usepackage[english]{babel}
\usepackage[latin1]{inputenc}

\usepackage{eucal}
\usepackage{esint}
\usepackage{amsgen}
\usepackage{amsthm}
\usepackage{xspace}
\usepackage{amssymb}
\usepackage{amsmath}
\usepackage{MnSymbol}
\usepackage{amsfonts}
\usepackage{verbatim}
\usepackage{mathrsfs, dsfont}

\usepackage{a4wide}

\usepackage{microtype}
\usepackage[bf, up, hang]{caption}

\usepackage{indentfirst}
\usepackage{graphicx}
\usepackage{subfigure}
\usepackage[section]{placeins}
\usepackage{psfrag}

\usepackage[usenames, dvipsnames, pdf]{pstricks}
\usepackage{epsfig}
\usepackage{pst-grad} 
\usepackage{pst-plot} 

\usepackage{color}
\definecolor{oneblue}{rgb}{0.0, 0.0, 0.85}
\definecolor{darkgrey}{rgb}{0.273, 0.281, 0.30}

\usepackage[colorlinks,
            urlcolor=oneblue,
            linkcolor=oneblue,
            citecolor=oneblue,
            bookmarksopen=false,
            pagebackref]{hyperref}

\usepackage[compact]{titlesec}

\titleformat{\section}{\normalfont\Large\bfseries\sffamily\center\color{darkgrey}}{\thesection.}{0.5em}{}{}
\titleformat{\subsection}{\normalfont\large\bfseries\sffamily\color{darkgrey}}{\thesubsection.}{0.4em}{}{}
\titleformat{\subsubsection}{\normalfont\normalsize\bfseries\sffamily\color{darkgrey}}{\thesubsubsection.}{0.3em}{}{}

\titlespacing*{\section}{1.0em}{1.0em}{0.8em}[0em]
\titlespacing*{\subsection}{1.0em}{1.0em}{0.8em}[0em]
\titlespacing*{\subsubsection}{1.0em}{0.7em}{0.6em}[0em]

\usepackage{titletoc}

\contentsmargin{0.0em}
\dottedcontents{section}[2.5em]{\addvspace{0.45em}\bfseries}{1.0em}{0pc}
\dottedcontents{subsection}[4.5em]{}{2.6em}{0pc}
\dottedcontents{subsubsection}[5.5em]{}{3.0em}{0pc}

\usepackage{fancyhdr}
\usepackage{lastpage}

\newcommand*\Title{Camassa--Holm equations for axisymmetric pipe flows}
\newcommand*\Authors{F.~Fedele \& D.~Dutykh}

\usepackage{acronym}

\numberwithin{equation}{section}

\newcommand{\ud}{\mathrm{d}}

\newcommand{\ue}{\mathrm{e}}
\newcommand{\N}{\mathcal{N}}
\newcommand{\Ls}{\mathsf{L}}

\renewcommand{\O}{\mathcal{O}}
\renewcommand{\L}{\mathcal{L}}
\renewcommand{\S}{\mathcal{S}}

\newcommand{\RE}{\mathrm{Re}}

\newcommand{\pd}[2]{\frac{\partial\, #1}{\partial\/ #2}}

\newcommand{\half}{{\textstyle{1\over2}}}

\begin{document}

\title[\Title]{Camassa--Holm type equations for axisymmetric Poiseuille pipe flows}

\author[F~.Fedele]{Francesco Fedele$^*$}
\address{School of Civil and Environmental Engineering \& School of Electrical and Computer Engineering, Georgia Institute of Technology, Atlanta, USA}
\email{fedele@gatech.edu}
\urladdr{http://www.ce.gatech.edu/people/faculty/511/overview}
\thanks{$^*$ Corresponding author}

\author[D.~Dutykh]{Denys Dutykh}
\address{University College Dublin, School of Mathematical Sciences, Belfield, Dublin 4, Ireland \and LAMA, UMR 5127 CNRS, Universit\'e de Savoie, Campus Scientifique, 73376 Le Bourget-du-Lac Cedex, France}
\email{Denys.Dutykh@ucd.ie}
\urladdr{http://www.denys-dutykh.com/}

\begin{abstract}
We present a study of the nonlinear dynamics of a disturbance to the laminar state in non-rotating axisymmetric Poiseuille pipe flows. The associated Navier-Stokes equations are reduced to a set of coupled generalized Camassa-Holm type equations. These support singular inviscid travelling waves with wedge-type singularities, the so called peakons, which bifurcate from smooth solitary waves as their celerity increase. In physical space they correspond to localized/periodic toroidal vortices or vortexons. The inviscid vortexon is similar to the nonlinear neutral structures  found by \textsc{Walton} (2011) \cite{Walton2011} and it may be a precursor to puffs and slugs observed at transition, since most likely it is unstable to non-axisymmetric disturbances.
\end{abstract}


\maketitle

\tableofcontents

\section{Introduction}

Transition to turbulence in non-rotating pipe flows is triggered by finite-amplitude perturbations \cite{Hof2003} since the laminar Hagen--Poiseuille flow is believed to be linearly stable to periodic or localized infinitesimal perturbations for all Reynolds numbers $\RE$ (see, for example, \cite{Drazin2004}). The coherent structures observed at the transitional stage are in the form of localized patches known as puffs and slug structures \cite{Wygnanski1973, Wygnanski1975}. Puffs are spots of vorticity localized near the pipe axis surrounded by laminar flow. The associated vorticity field is locally three-dimensional with a non negligible axisymmetric or toroiodal component as observed in experiments \cite{Wygnanski1973}. Slugs develop along the streamwise direction, while expanding through the entire cross-section of the pipe, and they are concentrated near the wall. Previous theoretical studies tried to relate slug flows to quasi inviscid solutions of the Navier--Stokes (NS) equations for non-rotating pipe flows. In particular, for non-axisymmetric pipe flows \textsc{Smith} \& \textsc{Bodonyi} (1982) \cite{Smith1982} revealed the existence of nonlinear neutral structures localized near the pipe axis (centre modes) in the form of inviscid travelling waves of small but finite amplitude, which are unstable equilibrium states (see \cite{Walton2005}). More recently, \textsc{Walton} (2011) \cite{Walton2011} found the axisymmetric analogue of Smith and Bodony's modes. Such inviscid axisymmetric structures are similar to the slugs of vorticity that have been observed in both experiments \cite{Wygnanski1973} and numerical simulations \cite{Willis2008a}. Thus, they may play a role in pipe flow transition as precursors to puffs and slugs.

Recently \textsc{Fedele} (2012) \cite{Fedele2012b} investigated the dynamics of non-rotating axisymmetric pipe flows in terms of travelling waves of nonlinear wave equations. He showed that, at high Reynolds numbers, the dynamics of small but finite long-wave perturbations of the laminar flow obey a coupled system of nonlinear Korteweg-de Vries-type (KdV) equations. These set of equations generalize the one-component KdV model derived by \textsc{Leibovich} \cite{Leibovich1968, Leibovich1969, Leibovich1970} to study propagation of waves along the core of concentrated vortex flows (see also \cite{Benney1966}) and vortex breakdown \cite{Leibovich1984}. \textsc{Fedele}'s coupled KdV equations support inviscid soliton and periodic wave solutions in the form of toroidal vortex tubes, hereafter referred to as \emph{vortexons}, which are similar to the inviscid nonlinear neutral centre modes found by \textsc{Walton} (2011) \cite{Walton2011}. Note that nonlinear dispersive wave equations arise in similar studies of the dynamics of Blasius flows, which at high Reynolds numbers is described by a Benjamin-Davis-Acrivos (BDA) integro-differential equation \cite{Ryzhov2010}. This supports soliton structures that explain the formation of spikes observed in boundary-layer transition \cite{Kachanov1993}.

In this paper, we extend the previous analysis in \cite{Fedele2012b} and show that the axisymmetric NS equations for non-rotating pipe flows can be reduced to a set of generalized coupled Camassa-Holm equations \cite{Camassa1993} that support inviscid traveling waves. Finally, the intepretation of the associated vortical structures is discussed.

\section{Camassa-Holm type equations for axisymmetric pipe flows}

Consider the axisymmetric motion of an incompressible fluid in a pipe of circular cross section of radius $R$ driven by an imposed uniform pressure gradient. Define a cylindrical coordinate system $(z, r, \theta)$ with the $z$-axis along the streamwise direction, and $(u, v, w)$ as the radial, azimuthal and streamwise velocity components. The time, radial and streamwise lengths as well as velocities are rescaled with $T$, $R$ and $U_0$ respectively. Here, $T = R/U_0$ is a convective time scale and $U_0$ is the maximum laminar flow velocity. A cylindrical divergence-free axisymmetric velocity field is given in terms of a Stokes streamfunction $\Psi(r,z,t)$ as
\begin{equation*}\label{eq:stream}
  u = -\frac{1}{r}\pd{\Psi}{z}, \qquad w = \frac{1}{r}\pd{\Psi}{r}.
\end{equation*}
To study the nonlinear dynamics of a perturbation superimposed on the laminar base flow $W_0(r) = 1-r^2$, $\Psi$ is decomposed as
\begin{equation}
  \Psi = \Psi_0 + \psi,
\end{equation}
where $\Psi_0 = \half r^2\bigl(1 - \half r^2\bigr)$ represents the stream function of the laminar flow $W_0$, and $\psi$ that of the disturbance. The curl of the NS equations yields the following nonlinear equation for $\psi$ \cite{Itoh1977}:
\begin{equation}\label{eq:fg}
  \partial_t\Ls\psi + W_0\partial_z\Ls\psi - \frac{1}{\RE}\Ls^2\psi = \N(\psi),
\end{equation}
where the nonlinear differential operator
\begin{equation*}
\N(\psi) = -r^{-1}\partial_r\psi\partial_z\Ls\psi + r^{-1}\partial_z\psi\partial_r\Ls\psi - 2r^{-2}\partial_z\psi\Ls\psi,
\end{equation*}
the linear operator
\begin{equation*}\label{eq:oper}
\Ls = \L + \partial_{zz}, \qquad \L = \partial_{rr} - r^{-1}\partial_r \equiv r\partial_r\left(r^{-1}\partial_r\right),
\end{equation*}
and $\RE$ is the Reynolds number based on $U_0$ and $R$. The boundary conditions for \eqref{eq:fg} reflect the boundedness of the flow at the centerline of the pipe and the no-slip condition at the wall, that is
\begin{equation*}
  \partial_r\psi = \partial_z\psi = 0 \quad \mbox{ at } r = 1.
\end{equation*}

Drawing from \cite{Fedele2012b}, the solution of \eqref{eq:fg} can be given in terms of a complete set of orthonormal basis $\{\phi_j(r)\}$ as 
\begin{equation}\label{eq:ex1}
  \psi(r,z,t) = \sum_{j=1}^J\phi_j(r) B_j(z,t),
\end{equation}
where $B_{j}$ is the amplitude of the radial eigenfunctions $\phi _{j}$ that satisfy the Boundary Value Problem (BVP) (see \cite{Fedele2005, Fedele2012b})
\begin{equation*}
  \L^2\phi_j = -\lambda_j^2\L\phi_j
\end{equation*}
with boundary conditions
\begin{eqnarray}\label{eq:bc1}
  \frac{1}{r}\phi_j < \infty, \quad r^{-1}\partial_r\phi_j < \infty \quad
\mbox{ as }& \quad r\rightarrow +0, \\
  \phi_j = \partial_r\phi_j = 0 \quad \mbox{ at }& \quad r = 1. \label{eq:bc2}
\end{eqnarray}
The positive eigenvalues $\lambda_j$ are the roots of $J_2(\lambda_j) = 0$, where $J_2(r)$ are the Bessel functions of the first kind of second order (see \cite{Abramowitz1965}). The corresponding eigenfunctions
\begin{equation*}
\phi_n = c_n\left[r^2 - \frac{rJ_1\left(\lambda_n r\right)}{J_1\left(\lambda_n\right)}\right],
\end{equation*}
form a complete and orthonormal set with respect to the inner product
\begin{equation*}
\left\langle\varphi_1, \varphi_2\right\rangle = -\int\limits_0^1\varphi
_1\;\L\varphi_2\frac{\ud r}{r} = \int\limits_0^1 \partial_r\varphi
_1\partial_r\varphi_2 \frac{\ud r}{r}.
\end{equation*}
For the first two least stable modes $\lambda_1 \approx 5.136$ and $\lambda_2 \approx 8.417$, respectively. Since $\phi_j$ satisfies the pipe flow boundary conditions \eqref{eq:bc1} and \eqref{eq:bc2} \emph{a priori}, so does $\psi$ of \eqref{eq:ex1}. A Galerkin projection of \eqref{eq:fg} onto the Hilbert space $\S$ spanned by $\{\phi_j\}_{j=1}^{N}$ yields a set of coupled generalized Camassa--Holm (CH) equations \cite{Camassa1993}
\begin{equation}\label{eq:main2}
\partial_t B_j + c_{jm}\partial_z B_m + \beta_{jm}\partial_{zzz}B_m + \alpha_{jm}\partial_{zzt}B_m + N_{jnm}(B_n, B_m) + \frac{\lambda_j^2}{\RE}B_j = 0, \quad j=1,\dots, N,
\end{equation}
where the nonlinear tensor operator 
\begin{equation*}
N_{jnm}(B_n, B_m) = F_{jnm}B_n\partial_z B_m + G_{jnm}\partial_z B_n\partial_{zz}B_m + H_{jnm}B_n\partial_{zzz}B_m.
\end{equation*}
The tensors $c_{jm},$ $\beta_{jm}$, $\alpha_{jm}$, $F_{jnm}$, $G_{jnm}$, $H_{jnm}$ are given in Appendix \ref{app:a} and summation over repeated indices is implicitly assumed. Note that CH type equations arise also as a regularized model of the 3-D NS equations (see \cite{Chen1999a, Domaradzki2001, Foias2001, Foias2002}), the so called Navier-Stokes-alpha model. Similarly to this, the truncated CH model \eqref{eq:main2} inhibits creation and excitation of smaller scales associated to higher damped modes $j > N$, since these are neglected.

\section{Singular Vortexons: CH Peakons}

\begin{figure}
  \centering
  \includegraphics[width=0.99\textwidth]{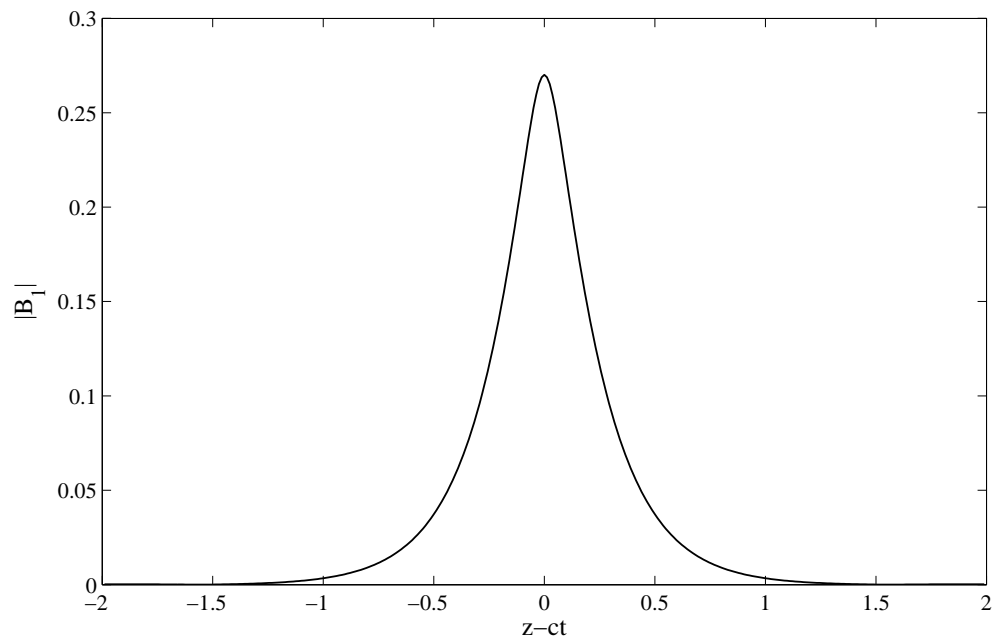}
  \caption{\em Regular solitary wave obtained numerically by the Petviashili method (dimensionless velocity $V = 0.60$).}
  \label{fig:fig1}
\end{figure}

\begin{figure}
  \centering
  \includegraphics[width=0.99\textwidth]{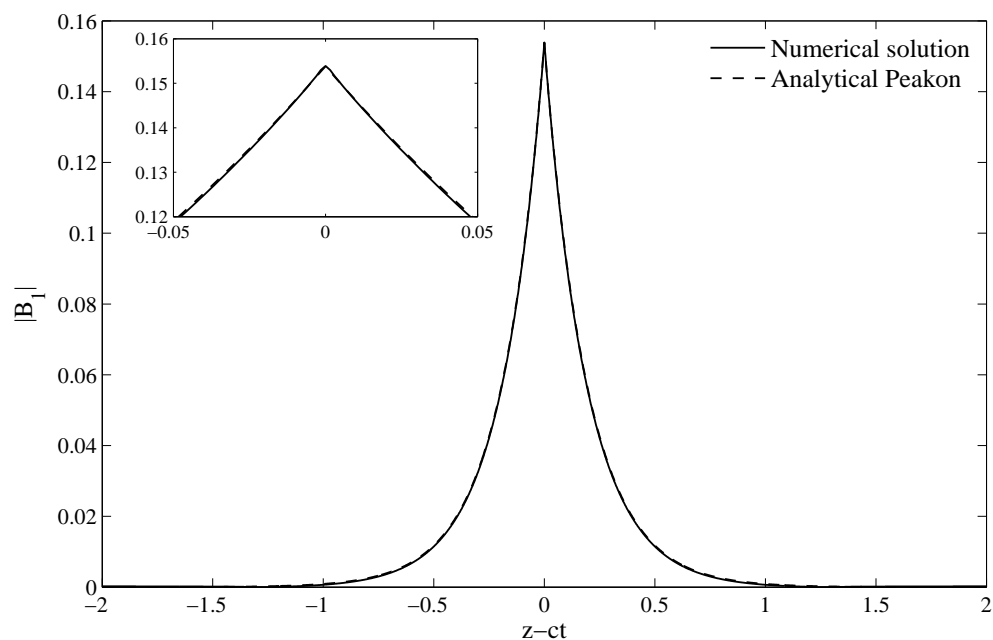}
  \caption{\em Analytical CH peakon (solid line) and numerical solution (dashed line) obtained by the Petviashili method (dimensionless velocity $V_1 \approx 0.63$).}
  \label{fig:fig2}
\end{figure}

\begin{figure}
  \centering
  \includegraphics[width=0.89\textwidth]{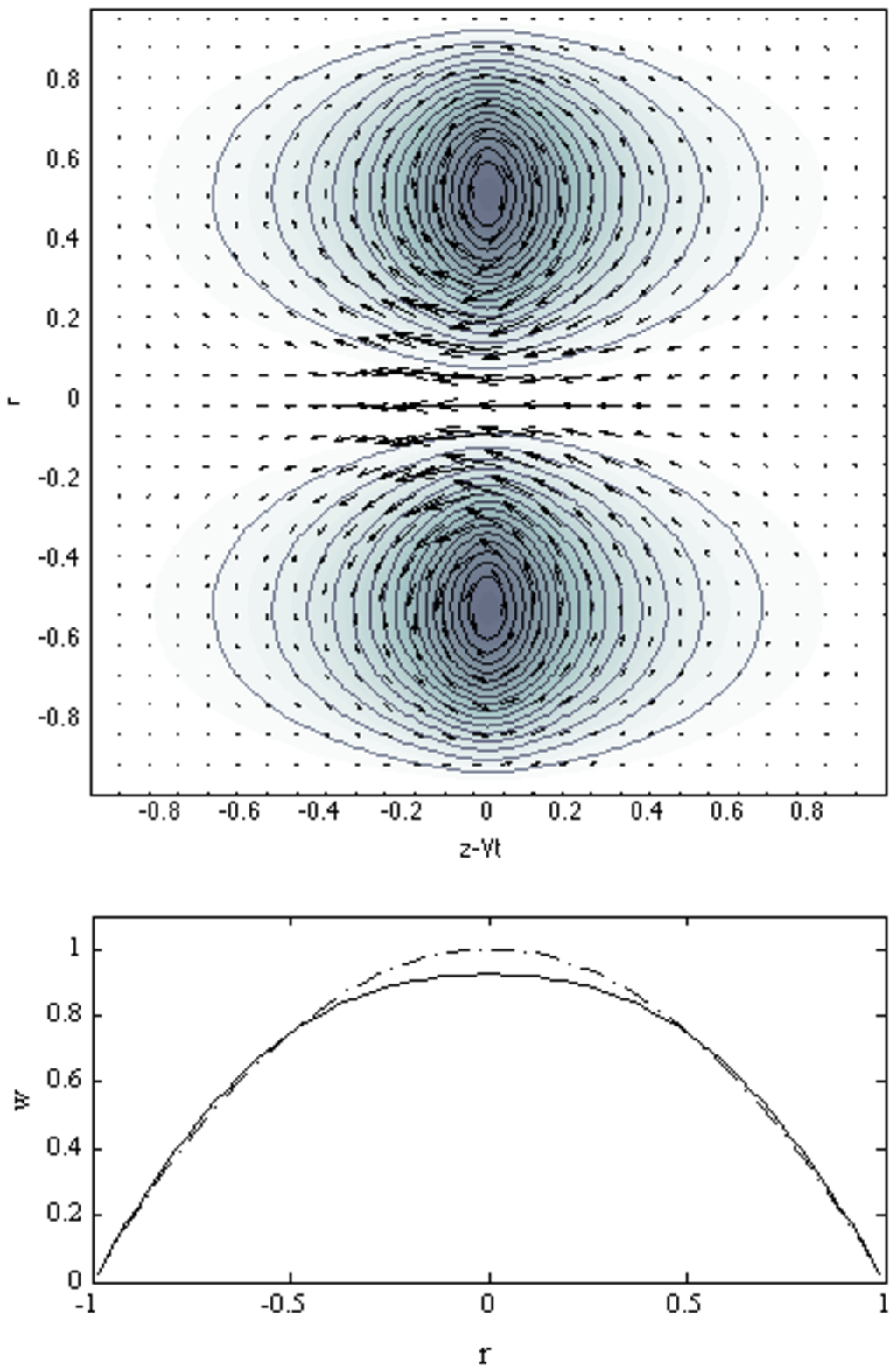}
  \caption{\em Regular vortexon: (top) streamlines of the perturbation and (bottom) velocity profiles of the perturbed (solid) and laminar (dash) flows.}
  \label{fig:fig3}
\end{figure}

\begin{figure}
  \centering
  \includegraphics[width=0.89\textwidth]{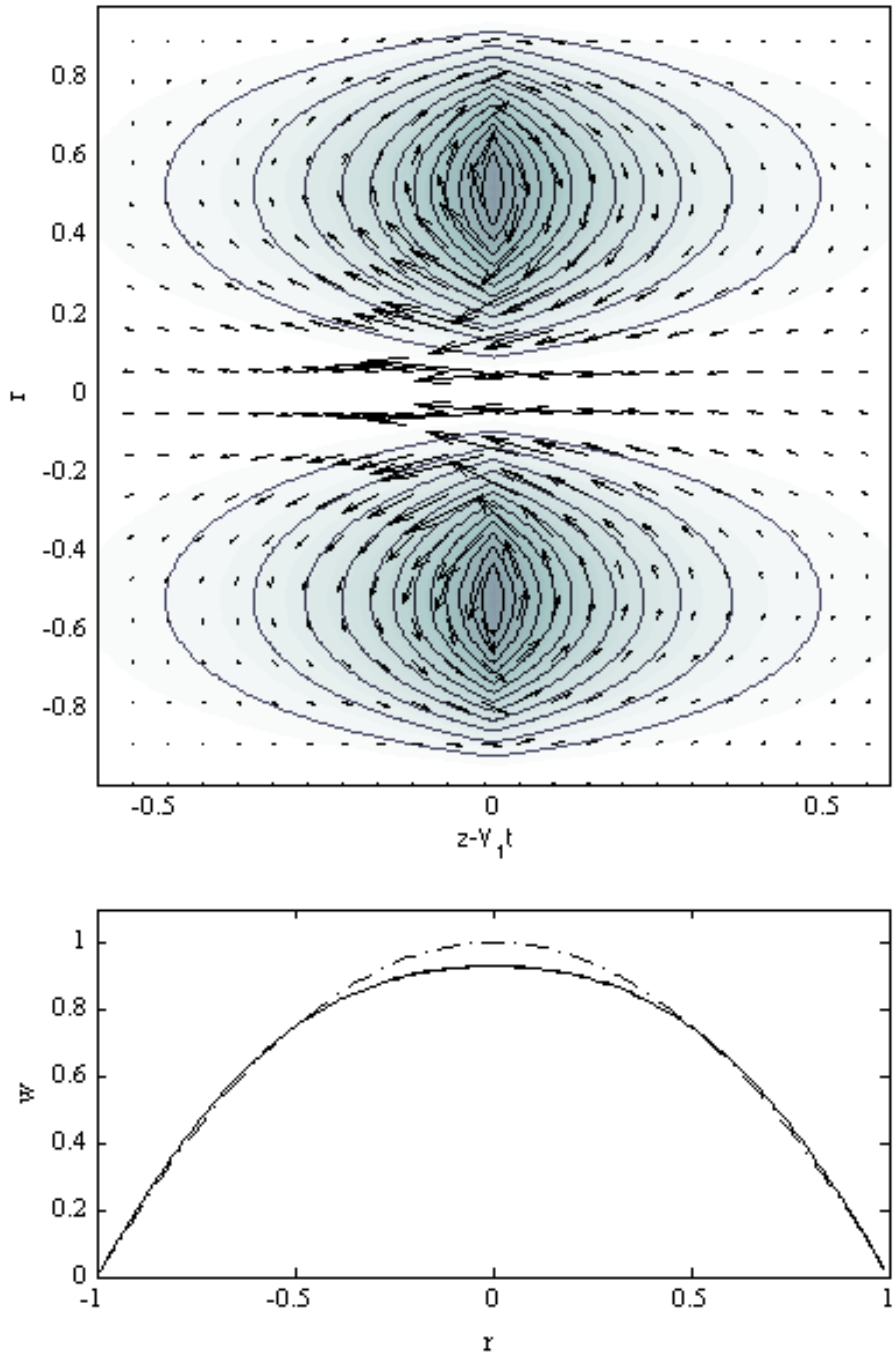}
  \caption{\em Singular vortexon: (top) streamlines of the perturbation and (bottom) velocity profiles of the perturbed (solid) and laminar (dash) flows.}
  \label{fig:fig4}
\end{figure}

Consider the inviscid version of the special case of the uncoupled CH equations
\begin{equation}\label{eq:CH2}
  \partial_{t}B_j + c_{jj}\partial_z B_j + \beta_{jj}\partial_{zzz}B_j + \alpha_{jj}\partial_{zzt}B_j + \N_j(B_j) = 0,
\end{equation}
where
\begin{equation*}
\N_j(B_j) = F_{jjj}B_j\partial_z B_j + G_{jjj}\partial_z B_j\partial_{zz} B_j + H_{jjj}B_j\partial_{zzz}B_j,
\end{equation*}
and no implicit summation over repeated indices. These support exponentially shaped singular solutions,the so called peakons, of the form
\begin{equation}\label{eq:peak}
  B_j(z,t) = a_j \ue^{-s_j\left\vert z-V_jt\right\vert},
\end{equation}
where
\begin{equation}\label{eq:coeff}
a_j = \frac{V_j\alpha_{jj} - \beta_{jj}}{H_{jjj}}, \qquad V_j = \frac{c_{jj} + \beta_{jj}s_j^2}{1 + \alpha_{jj} s_j^2}, \qquad s_j^2 = -\frac{F_{jjj}}{G_{jjj}+H_{jjj}}.
\end{equation}
Numerical computations revealed that ${s_j^2>0}$ and the peakon arises as a special balance between the linear dispersion terms $\partial_{zzz}B_j$, $\partial_{zzt}B_j$ and their nonlinear counterpart $B_j\partial_{zzz}B_j$ in \eqref{eq:CH2}. These three terms are interpreted in distributional sense because they give rise to Dirac delta functions that must vanish by properly chosing the amplitude $a_j$, thus satisfying the differential equation \eqref{eq:CH2} in the sense of distributions. The associated streamfunction $\psi_j^{(p)}$ is given by
\begin{equation*}\label{eq:streamsol}
  \psi_j^{(p)}(r,z,t) = a_j \ue^{-s_j^2\left\vert z-V_j t\right\vert}\phi_j(r).
\end{equation*}
The peakon \eqref{eq:peak} bifurcates from a regular solitary wave as the celerity increases above the dimensionless peakon speed $V_j$ in \eqref{eq:coeff} (normalized with respect to maximum laminar velocity $U_0$). For example, for the least stable eigenmode $B_1$ ($\lambda_1 \approx 5.136$), $V_1 \approx 0.63$. Figure~\ref{fig:fig1} shows a regular soliton at speed $V=0.60$ computed using the Petviashili method (see \cite{Pelinovsky2004, Lakoba2007, Fedele2011, Fedele2012a, Fedele2012}). A peakon bifurcates as the speed increases above $V_1$ and it is shown in Figure~\ref{fig:fig2}. The vortical structure (streamlines) of the perturbation associated to the regular and singular solitons are shown in the top panel of Figures~\ref{fig:fig3} and  \ref{fig:fig4}, respectively. These correspond to localized toroidal vortices that wrap around the pipe axis (centre vortexons). In particular, the vortexon associated to a peakon has discontinuous radial velocity $u$ across $z - ct = 0$ (see top panel of Figure \ref{fig:fig3}), but continuous streamwise velocity $w$ since the mass flux through the pipe is conserved. As a result, a sheet of azimuthal vorticity is advected at speed $V_1$. At the centre ($z-ct=0$) the profile of the streamwise velocity $W_0 + w$ of the perturbed flow (laminar base flow plus a vortexon) is shown in the bottom panels of Figures~\ref{fig:fig3} and \ref{fig:fig4} for the singular and regular vortexons respectively. Their effect is to slowdown the faster laminar flow near the core of the pipe by advecting the slower flow at the wall toward the pipe axis. Similar vortexons are also found numerically for the three-component CH equations \eqref{eq:main2} using the Petviashili method, but these results will be discussed elsewhere.

Finally we note that, as for the original CH equation \cite{Camassa1993}, viscous dissipation rules out the emergence of peakons and only smooth vortexons appear in the dynamics. As $\RE \to \infty$, a vortexon of amplitude $\sim \O(\RE^{-2.5})$ eventually decays due to viscous effects on the longer time scale $t \sim \O(\RE^{6.25})$ (see \cite{Fedele2012b}). As a result, under the axisymmetric dynamics the soliton structures can be assumed to behave inviscidly on shorter time scales. Thus, before they decay vortexons may be prone to instability due to non-axisymmetric perturbations (see, for example, \cite{Walton2005}).

\section{Conclusions}

We investigated the nonlinear dynamics of a disturbance to the laminar state in non-rotating axisymmetric Poiseuille pipe flows. The associated Navier-Stokes equations are projected onto the function space spanned by a finite set of the first few least stable Stokes eigenmodes. The eigenmode amplitudes depend upon both the streamwise direction and time and satisfy a truncated set of coupled generalized CH equations. For the uncoupled equations we found analytically special inviscid travelling waves with wedge-type singularities, \emph{viz}. peakons, which bifurcate from regular solitary waves as their celerity increase above a well defined threshold. In physical space peakons correspond to localized toroidal vortical structures with discontinuous radial velocities that wrap around the pipe axis (singular centre vortexons). Clearly, the inviscid singular vortexon could be an artifact of the Galerkin truncation of the axisymmetric Euler equations. However, it may be an approximation of singular solutions of the axisymmetric Euler equations (see, for example, \cite{Eyink2008}) and susceptible to Kelvin-Helmholtz type instability mechanisms. We point out that the inviscid centre vortexon is similar to the neutral mode identified by \textsc{Walton} (2011) \cite{Walton2011} and to the inviscid axisymmetric \emph{slug} structure proposed by \textsc{Smith} \emph{et al}. (1990) \cite{Smith1990}. They may play a role in pipe flow transition as precursors to puffs and slugs, since most likely they are prone to instability by non-axisymmetric disturbances (see \cite{Walton2005}).

\section*{Acknowledgements}
\addcontentsline{toc}{section}{Acknowledgments}

D.~\textsc{Dutykh} acknowledges the support from ERC under the research project ERC-2011-AdG 290562-MULTIWAVE. F.~\textsc{Fedele} acknowledges the travel support received by the Geophysical Fluid Dynamics (GFD) Program to attend part of the summer school on ``\emph{Spatially Localized Structures: Theory and Applications}'' at the Woods Hole Oceanographic Institution in August 2012.

\appendix

\section*{}\label{app:a}
\addcontentsline{toc}{section}{Appendix}

\begin{equation*}
  c_{jm} = -\int\limits_0^1 W_{0}\phi_{j}\L\phi_{m}\;r^{-1}\,\ud r,\quad 
  \alpha_{jm} = -\int\limits_0^1 \phi_j\phi_m\;r^{-1}\,\ud r, \quad 
  \beta_{jm} = -\int\limits_0^1 W_{0}\phi_j\phi_m r^{-1}\,\ud r,
\end{equation*}
\begin{equation*}
  F_{jnm} = -\int\limits_0^1\phi_j\left[\partial_r\phi_n\L\phi_m - \partial_r\left(\L\phi_n\right)\phi_m + 2r^{-1}\L\phi_n\phi_m\right]r^{-2}\,\ud r,
\end{equation*}
\begin{equation*}
  H_{jnm} = -\int\limits_0^1\phi_j\phi_m\partial_r\phi_n r^{-2}\,\ud r, \quad
  G_{jnm} = -\int\limits_0^1\phi_j\left[-\phi_m\partial_r\phi_n + 2r^{-1}\phi_n\phi_m\right]r^{-2}\,\ud r.
\end{equation*}

\bibliography{biblio}
\bibliographystyle{unsrt}
\addcontentsline{toc}{section}{References}

\end{document}